\documentclass[letterpaper,11pt]{amsart}
\usepackage{amssymb}
\usepackage{url}
\usepackage{pstricks}

\swapnumbers
\theoremstyle{plain}
\newtheorem{thm}{Theorem}[section]
\newtheorem{lem}[thm]{Lemma}
\newtheorem{prop}[thm]{Proposition}
\newtheorem{cor}[thm]{Corollary}

\newtheorem*{OpQu*}{Open question}
\theoremstyle{definition}
\newtheorem{defn}[thm]{Definition}
\theoremstyle{remark}
\newtheorem{note}[thm]{Note}
\newtheorem*{note*}{Note}
\newtheorem{exmp}[thm]{Examples}
\newtheorem*{exmp*}{Examples}

\numberwithin{equation}{thm}

\DeclareMathOperator{\Coef}{Coef}

\DeclareMathOperator{\XOR}{\scriptstyle{\mathsf{XOR}}}
\DeclareMathOperator{\OR}{\scriptstyle{\mathsf {OR}}}
\DeclareMathOperator{\AND}{\scriptstyle{\mathsf {AND}}}

\DeclareMathOperator{\NOT}{\scriptstyle{\mathsf {NOT}}}
\DeclareMathOperator{\SHL}{\scriptstyle{\mathsf {SHL}}}
\DeclareMathOperator{\SHR}{\scriptstyle{\mathsf {SHR}}}

\newcommand{\Z}{\mathbb Z}

\newcommand{\N}{\mathbb N}

\newcommand{\I}{\mathbb I}

\renewcommand{\:}{\colon}
\renewcommand{\>}{\rightarrow}

\usepackage[backref,
pagebackref,%
bookmarks=true,%
colorlinks=true%
]%
{hyperref}

\title {Wreath Products in Stream Cipher Design
}

\author{Vladimir Anashin}
\address{Faculty of Information Security, 
Russian State University for the Humanities,\\
Kirovogradskaya Str., 25/2, Moscow 113534, Russia}

\email{anashin@rsuh.ru, vladimir@anashin.msk.su, vs-anashin@yandex.ru}

\begin{document}
\begin{abstract}
The paper develops a novel approach to stream cipher design: Both the
state update function and the output function of the corresponding pseudorandom generators
are compositions of arithmetic and
bitwise logical operations, which are standard instructions of
modern microprocessors. Moreover, 
both the state
update function and  the output function are being modified 
dynamically during the encryption. 
Also, these 
compositions could be keyed, so the only
information available to an attacker is
that these functions 
belong to some exponentially large class.

The paper shows that under rather loose conditions
the output sequence is uniformly distributed,
achieves maximum period length and has high linear complexity and $\ell$-error
linear complexity. Ciphers of
this kind are
flexible:
One could choose a suitable combination of instructions to obtain
due performance without affecting the quality of the output sequence. Finally, some evidence
is given that a key recovery
problem for 
(reasonably designed) stream ciphers of this kind is intractable up to 
plausible conjectures.

\end{abstract}
\maketitle
\section {Introduction}
\label{Intro}

A classical stream cipher  is usually thought of as a pseudorandom generator which produces a keystream, that is, a binary random-looking string. Encryption
procedure is just a bitwise addition modulo 2 (also called XORing) of the keystream to
a
plaintext, which is  represented as a binary string either. That is, a pseudorandom
generator is an algorithm that takes a short random string ({\it a key},
or {\it a seed}) and expands it into a very long random-looking string, a
keystream. 

To make software implementations of these algorithms platform-independent
as well as to achieve high performance, the algorithms must use only those instructions
that are common for contemporary processors. These instructions are 
numerical operations (addition,
multiplication, subtraction,..) and 
logical ones (bitwise 
exclusive {\it or}, $\XOR$, bitwise {\it and}, $\AND$, etc.).

All these numerical and bitwise logical operations, and whence,
all 
their compositions, belong to a special class of mappings from $n$-bit
words into $n$-bit words: Each $i$\textsuperscript{th}
bit of the output word depends only on bits $0,1,\ldots,i$ of input words.\footnote{These
mappings are well-known mathematical objects (however, under different names: Compatible mappings in algebra, determined
functions in automata theory, triangle boolean mappings in the theory of Boolean
functions, 
functions that satisfy Lipschitz condition
with constant 1 in $p$-adic analysis) dating back
to 1960\textsuperscript{th} \cite{LN}, \cite{Yb}. Usefulness of these mappings in cryptography
has being directly pointed out since 1993 by V.S. Anashin \cite{me-conf},
\cite{me-1}, \cite{me-exp},
\cite{me-2}, \cite{me-Kol}, \cite{me-04}. The name "T-functions"
for these mappings was suggested by   A. Klimov and A.
Shamir in 2002 \cite{KlSh}.} 
This
fact underlies a number of results that enable one to determine whether a function of this kind is one-to-one,
i.e., induces a permutation on $n$-bit words, or  whether this permutation
is a single cycle, or whether the function is 
balanced;
that is, for each $n$-bit word the number of all its preimages is exactly
the same, etc. 
Systematical studies of these properties for the above mentioned mappings
were started by \cite{me-conf} and \cite{me-1} (see also \cite{me-exp})
followed by \cite{Kot},\cite{me-2},\cite{me-Kol}, \cite{me-04},\cite{me-04a}, as well as by later works \cite{KlSh},
\cite{KlSh-2}, and \cite{KlSh:3}. 

The main goal of the paper is to present a mathematical background for
a novel approach to the design
of stream ciphers.
\footnote{This approach has been already resulted in a very fast and flexible stream
cipher ABC v.2, see \cite{abc_per},\cite{abc-v2}.
} 
In this design, recurrence laws that define the key-stream are 
combinations of the above mentioned numerical and logical operations;
moreover, these laws are being dynamically modified during encryption. Nevertheless,
under minor restrictions we are able to prove that the key-stream has the longest (of possible) period, uniform
distribution, and high linear complexity as well as high  $\ell$-error linear
complexity and high 2-adic span.
To give an idea of how these algorithms look like, consider the following
illustrative
example.

Let $m\equiv3\pmod 4$, $3\le m\le\frac{2^n}{n}$.
Take $m$ {\slshape arbitrary} compositions $v_0(x),\ldots,v_{m-1}(x)$ of the above
mentioned machine instructions (addition, multiplication, $\XOR$, $\AND$,
etc.), then take another $m$ {\slshape arbitrary} compositions 
$w_0(x),\ldots,w_{m-1}(x)$ of this kind. Arrange two arrays $V$ and $W$
writing these $v_j(x)$ and $w_j(x)$ to memory in {\slshape arbitrary} order. 
Now choose an arbitrary $x_0\in\{0,1,\ldots 2^n-1\}$ as a seed. The generator
calculates the recurrence
sequence of states $x_{i+1}=(i\bmod m+x_i+4\cdot v_{i\bmod m}(x_i))\bmod 2^n$
and outputs the sequence 
$z_{i}=(1+\pi(x_i)+4\cdot w_{i\bmod m}(\pi(x_i)))\bmod 2^n$,
where $\pi$ is a bit order reverse permutation, which reads an $n$-bit
number
$z\in\{0,1,\ldots, 2^{n}-1\}$ in a reverse bit order; e.g., $\pi(0)=0, \pi(1)=2^{n-1},
\pi(2)=2^{n-2},\pi(3)=2^{n-2}+2^{n-1}$, etc. Then the sequence $\{x_i\}$
of $n$-bit numbers
is periodic; its shortest
period is of length $2^nm$, and each
number of $\{0,1,\ldots, 2^{n}-1\}$ occurs at the period exactly $m$ times.
Moreover, replacing each number $x_i$ in $\{x_i\}$ by an $n$-bit word that is a base-$2$
expansion of $x_i$, 
we obtain by concatenation of these $n$-bit words a binary counterpart
of the sequence $\{x_i\}$, i.e., a binary sequence $\{x_i\}^\prime$
with a period
of length
$2^nmn$. This period is random in the sense of \cite[Section 3.5, Definition Q1]{Knuth}
(see \eqref{eq:Q1} further);
each $k$-tuple ($0< k\le n$) occurs in this
sequence $\{x_i\}^\prime$ with  frequency\footnote{we count overlapping $k$-tuples either} $\frac{1}{2^k}$ exactly. 
The output sequence $\{z_i\}$ of numbers is also
periodic; its shortest period is of length $2^nm$; each number of $\{0,1,\ldots, 2^{n}-1\}$
occurs at the period exactly $m$ times. Finally, length of the
shortest period of 
every binary subsequence $\{\delta_s(z_i)\colon i=0,1,2,\ldots\}$ 
obtained by reading $s$\textsuperscript{th}
bit $\delta_s(z_i)$  
($0\le s\le n-1$) 
of each member of the sequence $\{z_i\}$
is a multiple of $2^n$; linear complexity
of this binary subsequence $\{\delta_s(z_i)\}$
(as well as linear complexity of binary counterparts $\{z_i\}^\prime$ and $\{x_i\}^\prime$)
exceeds $2^{n-1}$.

Ciphers of this kind are rather flexible. For instance, in the above
example one can take $m=2^k$ instead of odd $m\equiv 3\pmod 4$ and replace $i\bmod m$
in the definition of the state transition functions by an arbitrary $c_i\in\{0,1,\ldots,2^k-1\}$.
To guarantee the above declared properties both of the state sequence and of the output
sequence one must only demand that $c_0+c_1+\cdots+c_{m-1}\equiv 1\pmod 2$.
Moreover, one can take instead of $\pi$
an arbitrary permutation
of bits that takes the leftmost bit to the rightmost position (for instance,
a circular $1$-bit rotation towards higher order bits, which is also a standard instruction
in modern microprocessors). Also, one can replace the second $+$ in the
definition of the state transition and/or output functions with $\oplus$
(i.e., with $\XOR$), or take the third summand in the form $2\cdot(w(\pi(x)+1)-w(\pi(x)))$
(or $2\cdot(w(\pi(x)+1)+\NOT(w(\pi(x)))$) instead of $4\cdot w(\pi(x_i))$,
etc. Once again we emphasize that both $v$ and $w$ could be arbitrary
compositions of the above mentioned machine instructions (and derived ones);
e.g., in the  above example one might take\footnote{this
example is of no practical value; it serves only to illustrate how `crazy'
the
compositions
could be}
$$v(x)=\Biggl(1+2\cdot\frac{(x\AND (x^{2}+x^3))\OR x^4}{3 + 4\cdot(5+6x^5)^{x^6\XOR x^7}}\Biggr)^{7+\frac{8x^8}{9+10x^9}}$$   
We assume here and on that all the operands are non-negative integer rationals
represented in their  base-2 expansions; so, for instance,  
$2=1\XOR 3 = 2\AND 7\equiv\NOT 13\pmod 8$, $\frac{1}{3}\equiv 3^{-1}\equiv
11\equiv -5\pmod {16}$, $3^{-\frac{1}{3}}\equiv
3^{11}\equiv 3^{-5}\equiv 11\pmod{16}$, etc.
Up to this agreement the functions $v$ and $w$ are well defined. 
The performance of the whole scheme depends only on the ratio of `fast' and `slow' operations 
in these compositions;
one may vary this ratio in a wide range to achieve desirable speed.

The paper is organized as follows. Section \ref{Prelm} concerns basic facts
about functions we use as `building blocks' of our generators, Section \ref{sec:Constr} describes
how to construct a generator out of these blocks, Section \ref{Out} studies
properties of output sequences of these generators, and Section \ref{Sec}
gives some reasoning why (some of) these generators could be provably secure.
Due to the space constraints, no proofs are given. 

\section{Preliminaries}
\label{Prelm}    

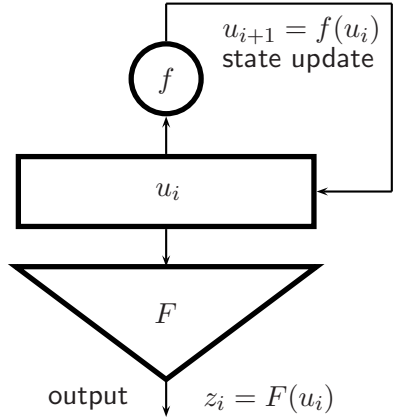
\begin{figure}
\begin{quote}\psset{unit=0.5cm}
 \begin{pspicture}(-2,0)(24,12)
\pscircle[linewidth=2pt](12,10){1}
  \psline(12,11)(12,12)
  \psline{->}(18,7)(16,7)
  \psline(18,12)(18,7)
  \psline(12,12)(18,12)
  \psline{->}(12,8)(12,9)
  \psline{->}(12,2)(12,1)
  \psline{<-}(12,5)(12,6)
  \psframe[linewidth=2pt](8,6)(16,8)
  \pspolygon[linewidth=2pt](8,5)(16,5)(12,2)
  \uput{0}[180](12.4,7){$u_i$}
  \uput{0}[90](12,9.6){$f$}
  \uput{1}[90](12,2.5){$F$}
  \uput{1}[0](12.5,11.3){$u_{i+1}=f(u_i)$}
  \uput{1}[0](12.5,10.5){{\sf state update}}
  \uput{1}[0](12,1.5){$z_i=F(u_i)$}
  \uput{1}[180](12,1.5){{\sf output}}
 \end{pspicture}
\end{quote}
\caption{Ordinary PRNG}
\label{fig:PRNG}
\end{figure}

Basically, the generator we consider in the paper is a finite automaton
${\mathfrak A}=\langle N,M,f,F,u_0\rangle $ with a finite state set $N$, state
transition function
$f:N\rightarrow N$, finite output alphabet $M$, output function  $F:N\rightarrow M$
and an initial state (seed) $u_0\in N$. Thus, this generator (see Figure
\ref{fig:PRNG}) produces a sequence
$$\mathcal S=\{F(u_0), F(f(u_0)), F(f^{(2)}(u_0)),\ldots, F(f^{(j)}(u_0)),\ldots\}$$ 
over
the set $M$, where 
$$f^{(j)}(u_0)=\underbrace{f(\ldots f(}_{j
\;\text{times}}u_0)\ldots)\ \ (j=1,2,\ldots);\quad f^{(0)}(u_0)=u_0.$$ 

\begin{figure}
\begin{quote}\psset{unit=0.5cm}
 \begin{pspicture}(-2,0)(24,12)
\pscircle[linewidth=2pt](12,10){1}
  \psline(12,11)(12,12)
  \psline{->}(18,7)(16,7)
  \psline(18,12)(18,7)
  \psline(12,12)(18,12)
  \psline{->}(12,8)(12,9)
  \psline{->}(12,2)(12,1)
  \psline{<-}(12,5)(12,6)
  \psframe[linewidth=2pt](8,6)(16,8)
  \pspolygon[linewidth=2pt](8,5)(16,5)(12,2)
  \uput{0}[180](12.4,7){$u_i$}
  \uput{0}[90](12,9.7){$f_i$}
  \uput{1}[90](12.1,2.3){$F_i$}
  \uput{1}[0](12.5,11.3){$u_{i+1}=f_i(u_i)$}
  \uput{1}[0](12.5,10.5){{\sf state update}}
  \uput{1}[0](12,1.5){$z_i=F_i(x_i)$}
  \uput{1}[180](12,1.5){{\sf output}}
 \end{pspicture}
\end{quote}
\caption{Counter-dependent PRNG}
\label{fig:cntdpd}
\end{figure}
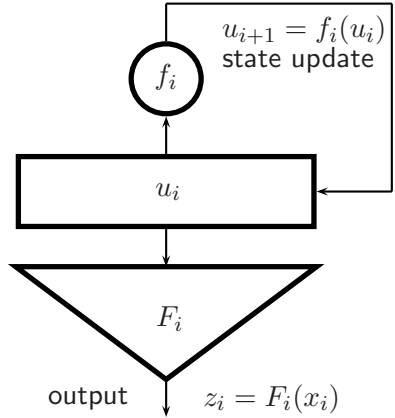

Automata of the form $\mathfrak
A$ could be used either as pseudorandom generators per se, or as components
of more complicated pseudorandom generators, 
the so called \textit {counter-dependent generators} (see Figure \ref{fig:cntdpd}); the latter produce 
sequences 
$\{z_0,z_1,z_2,\ldots\}$ over $M$ according to the rule
\begin{equation}
\label{eq:cntdpd}
z_0=F_0(u_0),u_1=f_0(u_0);\ldots
z_{i}=F_i(u_i), u_{i+1}=f_i(u_i);\ldots
\end{equation}
That is, at the $(i+1)$\textsuperscript{th} step the automaton 
$\mathfrak A_i=\langle N,M,f_i,F_i,u_i\rangle $
is applied to the state $u_i\in N$, producing a new state $u_{i+1}=f_i(u_i)\in
N$, and
outputting a symbol $z_{i}=F_i(u_i)\in M$.

Now we give a more formal
\begin{defn}
\label{def:WP}
Let $\mathfrak A_j=\langle N,M,f_j,F_j\rangle$  be a family of 
automata with the same state set $N$ and the same
output alphabet $M$ indexed by elements of
a non-empty (possibly, countably infinite) set $J$ 
(members of the family need not be necessarily pairwise distinct). 
Let $T\colon J\rightarrow J$ be an arbitrary mapping. A {\it wreath product}
 of the family $\{\mathfrak A_j\}$ of automata
with respect to the mapping $T$ is an automaton with the state set $N\times J$, state
transition function $\breve f(j,z)=(f_j(z),T(j))$ and output function 
$\breve F(j,z)=F_j(z)$. The state transition function $\breve f(j,z)=(f_j(z),T(j))$
is called a {\it wreath product of a family of mappings $\{f_j\colon j\in
J\}$ with respect to the mapping $T$} \footnote{cf. {\it skew shift} in ergodic theory;
cf. round function in the Feistel network. We are using a term from group theory.}. 
We call $f_j$ (resp., $F_j$) {\it clock} state update (resp.,
output) functions.
\end{defn}

It worth notice here  that if $J=\mathbb N_0$ and $F_i$ does not depend on $i$, this construction 
gives us a number of examples of counter-dependent generators 
in the sense of \cite[Definition 2.4]{ShTs}, where
the notion of a counter-dependent generator was 
originally introduced. 
However, we use this notion in a broader  sense
in comparison with that of \cite{ShTs}: In our counter-dependent
generators
not only the state
transition function, but also the output function depends on $i$. Moreover,
in \cite{ShTs} only a special case of counter-dependent generators
is studied; namely, counter-assisted generators and their cascaded and two-step
modifications. A state transition function of a counter-assisted generator  is
of the form $f_i(x)=i\star
h(x)$, where $\star$ is a binary quasigroup operation (in particular, group
operation, e.g., $+$ or
$\XOR$), and $h(x)$ does not depend on $i$. 
An output function of a counter-assisted generator does not depend
on $i$ either. Finally, our constructions provide long period, uniform
distribution, and high linear complexity of output sequences; cf. \cite{ShTs}, where only the
diversity is guaranteed.

Throughout the paper we assume 
that  $N=\mathbb I_n(p)=
\{0,1,\ldots,p^n-1\}$, $M=\mathbb I_m(p)$,  $m\le n$,  where $p$ is a 
prime. 
Moreover, mainly we are focused on the case $p=2$ as the
most suitable for computer implementations.
It is convenient to think of elements $z\in\I_n(p)$ as base-$p$ expansions
of rational integers: 
$$z=\delta_0^p(z)+\delta_1^p(z)\cdot p+\dots+\delta_{n-1}^p(z)\cdot
p^{n-1};$$ 
here $\delta_j^p(z)\in\{0,1,\dots,p-1\}$.
For $p=2$
we usually 
omit the superscript, when this does not lead to misunderstanding. Further
we usually identify $\mathbb I_n(p)$ with the ring $\Z/p^n$ of residues
modulo $p^n$.  

As said above,
we consider bitwise logical operators  
as functions defined on the set $\mathbb N_0=\{0,1,2,\ldots\}$
Machine instructions $\SHR_m$ and $\SHL_m$ --- 
an $m$-bit right shift ($\cdot \Rsh m$, which is a multiplication by $2^m$) and an $m$-bit
left shift ($\cdot\Lsh m$, integer division by $2^m$, i.e., $\lfloor\frac{\cdot}{2^m}\rfloor$,
with $\lfloor\alpha\rfloor$ being the greatest rational integer that does
not exceed $\alpha$) are defined on $\N_0$ either.
{\slshape Note that since this moment
throughout the paper we represent integers $i$
in reverse bit order ---
less significant bits left, according to their occurrences in $2$-adic
canonical representation
of $i=\delta_0(i)+\delta_1(i)\cdot 2+\delta_2(i)\cdot 4+\ldots$}; so
$0011$ is $12$, and not $3$. 
Moreover, one may think about these logical and machine operators, 
as well as of numerical,  i.e., arithmetic  ones (addition, multiplication, 
etc.),
as of functions that are
defined on  (and valuated in) the set $\mathbb Z_2$ of all $2$-adic integers\footnote{The latter 
ones within the context of this paper could be thought
of as countable infinite binary sequences with members indexed by $0,1,2,\ldots$;
$\Z_2$ is a metric space with respect to the $2$-adic norm $\|\alpha\|_2=2^{-k}$,
where $k$ is the number of the first zero members of the sequence $\alpha\in\Z_2$:
$\|0\|=\|000\ldots\|_2=0$, $\|1\|=\|100\ldots\|_2=1$, 
$\|2\|=\|010\ldots\|_2=\frac{1}{2}$, etc.}
(see \cite{me-1,
me-2}),
e.g., $x\OR y=(\delta_0(x)\vee\delta_0(y))+(\delta_1(x)\vee\delta_1(y))\cdot
2+(\delta_2(x)\vee\delta_2(y))\cdot 2^2+\ldots$.

A common feature  of the above mentioned 
operations 
is that they all, with exception of shifts towards less significant
bits and circular rotations\footnote{nevertheless, the both are used in
further constructions}, are {\it compatible}, i.e., $\omega(u,v)\equiv\omega(u_1,v_1)\pmod{2^r}$
whenever both congruences $u\equiv u_1\pmod{2^r}$  and $v\equiv v_1\pmod{2^r}$ hold
simultaneously. The notion of compatible mapping could be naturally generalized
to multivariate mappings $(\mathbb Z/p^l)^{t}\rightarrow(\mathbb Z/p^l)^{s}$ and 
$(\mathbb Z_p)^{t}\rightarrow(\mathbb Z_p)^{s}$ over a residue ring modulo
$p^l$ (resp., the ring $\Z_p$ of $p$-adic integers). 
Obviously, a composition of compatible mappings
is a compatible mapping. We list now some important examples of compatible
operators $(\mathbb Z_p)^{2}\rightarrow\mathbb Z_p$, $p$ prime (see
\cite{me-2}). Part of them originates from arithmetic operations:

\begin{equation}
\begin{split}
& {\text {\rm multiplication,}}\ \cdot:\ (u,v)\mapsto uv;\\ 
& {\text {\rm addition,}}\ +:\ (u,v)\mapsto u+v; 
\\
& {\text {\rm subtraction,}}\ -:\ (u,v)\mapsto u-v;
\\
& {\text {\rm exponentiation,}}\ \uparrow_p:\ (u,v)\mapsto u\uparrow_p v=(1+pu)^v;
\ {\text{\rm in particular,}}
\\ 
& {\text {\rm raising to negative powers}},\ u\uparrow_p(-r)=(1+pu)^{-r}, r\in\mathbb N;
\ {\text{\rm and}}
\\ 
& {\text {\rm division,}}\ /_p: u/_pv=u\cdot (v\uparrow_p(-1))=\frac{u}{1+pv}. 
\label{eq:opAr}
\end{split}
\end{equation}

The other part originates from digitwise logical operations of $p$-valued logic:
\begin{equation}
\label{eq:opLog} 
\begin{split}
& {\text {\rm digitwise multiplication}}\ u\odot_p v: \delta_j(u\odot_p
v)\equiv \delta_j (u)\delta_j (v)\pmod p;\\ 
& {\text {\rm digitwise addition}}\ 
u\oplus_p v: \delta_j(u\oplus_p
v)\equiv \delta_j (u)+\delta_j (v)\pmod p;\\ 
& {\text {\rm digitwise subtraction}}\
u\ominus_p v: \delta_j(u\ominus_p
v)\equiv \delta_j (u)-\delta_j (v)\pmod p.
\end{split}
\end{equation}
Here 
$\delta_j(z)$ $( j=0,1,2,\ldots)$
stands for the $j$\textsuperscript{th} digit of $z$ in its base-$p$ expansion.

More compatible mappings could be derived from the above mentioned
ones. For instance, a reduction modulo $p^n$, $n\in\mathbb N$, is $u\bmod p^n= u\odot_p
\frac{p^n-1}{p-1}$, an $l$-step shift towards more significant digits is just
a multiplication by $p^l$, etc. Obviously, $u\odot_2 v=u\AND v$, $u\oplus_2
v=u\XOR v$. Further in case $p=2$ we omit subscripts of the corresponding
operators. 

In case $p=2$ compatible mappings could be characterized in terms of Boolean
functions. Namely, each mapping $T\colon\mathbb Z/2^n\rightarrow\mathbb Z/2^n$ 
could be
considered as an ensemble of $n$ Boolean functions 
$\tau_i^T$,
$i=0,1,2,\ldots,n-1$,
in $n$ Boolean variables $\chi_0,\ldots,\chi_{n-1}$ by assuming $\chi_i=\delta_i(u)$,
$\tau_i^T(\chi_0,\ldots,\chi_{n-1})=\delta_i(T(u))$
for $u$ running from $0$ to $2^n-1$. The following proposition
holds.
\begin{prop}
\label{Bool} 
{\rm (\cite[Proposition 3.9]{me-1})}
A mapping $T\colon\mathbb Z/2^n\rightarrow\mathbb Z/2^n$ 
{\rm (}resp., a mapping $T\colon\mathbb Z_2\rightarrow\mathbb Z_2${\rm
)}
is compatible
iff each Boolean function $\tau_i^T(\chi_0,\chi_{1},\ldots)=\delta_i(T(u))$,
$i=0,1,2,\ldots$,
does not depend on the variables $\chi_{j}=\delta_j(u)$ for $j>i$.
\end{prop}
\begin{note*}
Mappings satisfying conditions of the proposition are also known
in the theory of Boolean functions as {\it triangle} mappings; the term
{\it $T$-functions} is used
in \cite{KlSh}, \cite{KlSh-2}, \cite{KlSh:3} instead. 
For multivariate mappings theorem \ref{Bool} holds either:
A mapping 
$T=(t_1,\ldots,t_s)\colon\mathbb (Z_2)^{(r)}\rightarrow\mathbb 
(Z_2)^{(s)}$ 
is compatible
iff each Boolean function $\tau_i^{t_j}(\chi_{1,0},\chi_{1,1},\ldots,
\chi_{r,0},\chi_{r,1},\ldots)=\delta_i(t_k(u,\ldots,u_r))$ ($i\in\N_0$,
$k=0,1,\ldots,s$) does not depend on the variables $\chi_{\ell,j}=\delta_j(u_{\ell})$ 
for $j>i$ ($\ell=1,2,\ldots,r$).
\end{note*}

Now, given a compatible mapping $T\colon\mathbb Z_2\rightarrow\mathbb Z_2$, one
can define an
induced mapping 
$T\bmod2^n\colon\mathbb Z/2^n\rightarrow\mathbb Z/2^n$
assuming $(T\bmod 2^n)(z) =T(z)\bmod 2^n=(T(z))\AND(2^n-1)$ 
for $z=0,1,\ldots,2^n-1$. Obviously, $T\bmod2^n$ is also compatible.
For odd prime
$p$, as well as for multivariate case 
$T\colon(\mathbb Z_p)^{s}\rightarrow(\mathbb Z_p)^{t}$
an induced mapping $T\bmod p^n$ could be defined by analogy.
\begin{defn}
\label{def:erg}
(See \cite{me-2}). We call a compatible mapping $T\colon\mathbb Z_p\rightarrow\mathbb Z_p$
{\it bijective modulo $p^n$} iff the induced mapping $T\bmod p^n$ is a permutation
on $\mathbb Z/p^n$; we call $T$ {\it transitive modulo $p^n$}, iff $T\bmod p^n$
is a  permutation with a single cycle. We say that $T$ is {\it
measure-preserving}
(respectively, {\it ergodic}),
iff $T$ is bijective (respectively, transitive) modulo $p^n$ for all $n\in\mathbb N$.
We call a compatible mapping
$T\colon(\mathbb Z_p)^{s}\rightarrow(\mathbb Z_p)^{t}$
{\it balanced modulo $p^n$} iff the induced mapping $T\bmod p^n$ maps
$(\mathbb Z/p^n)^{s}$ onto $(\mathbb Z/p^n)^{t}$, and each element of 
$(\mathbb Z/p^n)^{t}$ has the same number of preimages in $(\mathbb Z/p^n)^{s}$.
Also, the mapping $T\colon(\mathbb Z_p)^{s}\rightarrow(\mathbb Z_p)^{t}$ is 
called
{\it measure-preserving} iff it is balanced modulo $p^n$ for all $n\in\mathbb
N$.\footnote{
The terms measure-preserving and ergodic 
originate
from the theory of dynamical systems. Namely, a 
mapping 
$T\colon\mathbb Z_p\rightarrow\mathbb Z_p$ is compatible iff it satisfies
Lipschitz condition with a constant $1$ with respect to the $p$-adic metric;
$T$ defines a dynamics on the measurable
space
$\mathbb Z_p$ with respect to the 
normalized Haar measure.
The mapping $T$ is, e.g., ergodic with respect to this measure (in the
sense of the theory of dynamical systems) iff it satisfies \ref{def:erg},
see \cite{me-2} for details.
} 
\end{defn}
Both transitive modulo $p^n$ and balanced modulo $p^n$ mappings could
be used as building blocks of pseudorandom generators to provide both long
period
and uniform distribution of output sequences. The following obvious
proposition holds.
\begin{prop}
\label{prop:Auto}
If the state transition function $f$ of the automaton $\mathfrak A$ is
transitive on the state set $N$, i.e., if $f$ is a permutation with a single cycle
of length $|N|$; if, further, $|M|$ is a factor of $|N|$, and if the output function 
$F:N\rightarrow M$ is balanced
{\rm (}i.e.,  $|F^{-1}(s)|=|F^{-1}(t)|$ for all $s,t\in M${\rm )}, or,
in particular, bijective, then the output sequence
$\mathcal S$ of the automaton $\mathfrak A$ is purely periodic with a period of
length  $|N|$ 
{\rm (i.e., maximum possible)}, and each element of  
$M$ occurs at the period the same number of times: $\frac{|N|}{|M|}$ exactly. {\rm
That
is, the
output sequence $\mathcal S$ is uniformly distributed.} 
\end{prop}
\begin{defn}
\label{def:strict}
Further in the paper we call a sequence $\mathcal S=\{s_i\in M\}$ over a finite set
$M$ purely periodic with a period of length $t$ iff $s_{i+t}=s_i$ for all
$i=0,1,2,\ldots$. The sequence $\mathcal S$ is called {\it strictly uniformly
distributed} iff it is purely periodic with a period of length $t$, 
and every element of $M$ occurs at the period the same number of times,
i.e., exactly $\frac{t}{|M|}$. A sequence $\{s_i\in \mathbb Z_p\}$ of $p$-adic
integers is called {\it strictly uniformly
distributed modulo $p^k$} iff the sequence $\{s_i\bmod p^k\}$ of residues
modulo $p^k$ is strictly uniformly distributed over a residue ring $\mathbb Z/p^k$.
\end{defn}
\begin{note*}  A sequence $\{s_i\in \mathbb Z_p\colon i=0,1,2,\ldots\}$ of $p$-adic
integers is uniformly distributed (with respect to the normalized Haar measure
$\mu$   on $\mathbb
Z_p$) \footnote{i.e., $\mu(a+p^k\Z_p)=p^{-k}$ for all $a\in\Z_p$ and all
$k=0,1,2.\ldots$} iff it is uniformly distributed modulo $p^k$ for all $k=1,2,\ldots$;
that is, for every $a\in\mathbb Z/p^k$ relative numbers of occurrences 
of $a$ in the initial segment of length $\ell$ in the sequence 
$\{s_i\bmod p^k\}$ of residues
modulo $p^k$ 
are asymptotically equal,
i.e., 
$\lim_{\ell\to\infty}\frac{A(a,\ell)}{\ell}=\frac{1}{p^k}$, where 
$A(a,\ell)=|\{s_i\equiv a\pmod{p^k}\colon i<\ell\}|$ (see \cite{KN} for
details). So strictly uniformly distributed sequences are uniformly distributed
in the common meaning of the theory of distribution of sequences.
\end{note*}

Thus,
assuming $N=\mathbb Z/2^n, M=\mathbb Z/2^m, n=km$, 
$f=\overline f=\widetilde f\bmod {2^n}$ and $F=\overline
F=
\widetilde F\bmod{2^m}$,
where the function  $\widetilde f:\mathbb Z_2\rightarrow \mathbb Z_2$ is
compatible and ergodic, and the function 
$\widetilde F:(\mathbb Z_2)^{k}\rightarrow \mathbb Z_2$ is compatible
and measure-preserving, we obtain an automaton that generates a uniformly
distributed periodic sequence, and length of a  period of this sequence
is $2^n$. 
That is, each
element of $\mathbb Z/2^m$ occurs at the period the same number of times
(namely,
$2^{n-m}$). Obviously, the conclusion holds if one takes as 
 $F$ an arbitrary composition of the function
$\overline F=\widetilde F\bmod{2^m}$ with a measure-preserving function: For
instance, one may put $F(i)=\overline F(\pi(i))$ or $F(i)=\delta_j(i)$,
etc. 
Thus, proposition \ref{prop:Auto} makes it possible
to vary both the state transition and the output functions (for instance, to
make them key-dependent, or in order to achieve better performance\footnote{e.g.,
in \cite{KlSh} there was introduced a fast generator of this
kind: $f(x)=(x+(x^2\OR C))\bmod 2^{2n}$, $F(x)=\lfloor\frac{x}{2^n}\rfloor\bmod
2^n$}) {\slshape leaving the output sequence uniformly distributed}.

There exists an easy way to construct a measure preserving or ergodic
mapping out of an arbitrary compatible mapping, i.e., out of an arbitrary
composition of both arithmetic \eqref{eq:opAr} and logical \eqref{eq:opLog}
operators. 
\begin{prop}
\label{Delta} \cite[Lemma 2.1 and Theorem 2.5]{me-2}. Let $\Delta$ be a difference operator, i.e., $\Delta g(x)=g(x+1)-g(x)$
by the definition. Let, further, $p$ be a prime, let $c$ be a coprime with
$p$, $\gcd(c,p)=1$, and let $g\colon\mathbb Z_p\rightarrow
\mathbb Z_p$ be a compatible mapping. Then the mapping $z\mapsto c+z+p\cdot\Delta
g(z)\ (z\in\mathbb Z_p)$ is ergodic, and the mapping $z\mapsto d+cx+p\cdot
g(x)$
preserves measure for an arbitrary $d$.
Moreover, if $p=2$, then the converse also holds: Each compatible and ergodic
\textup {(}respectively, each compatible
and measure preserving\textup {)}
mapping $z\mapsto f(z)\ (z\in\mathbb Z_2)$ could be represented as
$f(x)=1+x+2\cdot\Delta g(x)$  \textup {(}respectively, as
$f(x)=d+x+2\cdot g(x)$\textup {)} for suitable $d\in\mathbb Z_2$ and compatible 
$g\colon\mathbb
Z_2\rightarrow \mathbb Z_2$.
\end{prop}
\begin{cor}
\label{erg-comp} 
Let $p=2$,
and let $f$
be a compatible
and ergodic mapping of $\mathbb Z_2$ onto itself. Then for each $n=1,2,\ldots$
the state transition function $f\bmod 2^n$ could be represented as a finite
composition of bitwise logical and arithmetic operators.
\end{cor}
For the sequel we
need one more representation, in a Boolean form (see \ref{Bool}). 
The following theorem is just a restatement of a known result from the
theory 
of Boolean functions, the so-called bijectivity/transitivity criterion  for triangle
Boolean mappings.
However, the criterion belongs to the mathematical folklore; thus it is 
difficult to
attribute it to somebody, yet a reader could find a proof in, e.g.,
\cite[Lemma 4.8]{me-1}. Recall that every Boolean function $\psi(\chi_0,\ldots\chi_n)$
in the Boolean variables $\chi_0,\ldots\chi_n$
admits a unique representation in the form
$$\psi(\chi_0,\ldots\chi_n)\equiv\sum_{\varepsilon_0,\ldots,\varepsilon_n\in\{0,1\}}
\xi_{\varepsilon_0,\ldots,\varepsilon_n}\chi_0^{\varepsilon_0}\cdots\chi_i^{\varepsilon_n}\pmod
2,$$
where $\xi_{\varepsilon_0,\ldots,\varepsilon_n}\in\{0,1\}$; the sum in
the right hand part is called
an algebraic normal
form (ANF) of the Boolean function $\psi$. The degree $\deg \psi$ is $\max\{\varepsilon_0+\cdots+\varepsilon_n\colon
\xi_{\varepsilon_0,\ldots,\varepsilon_n}=1\}$.
\begin{thm}
\label{ergBool} 
A mapping $T\colon\mathbb Z_2\rightarrow\mathbb Z_2$ is
compatible and measure-preserving iff for each $i=0,1,\ldots$ the ANF of
the Boolean function 
$\tau^T_i=\delta_i(T)$
in Boolean variables $\chi_0,\ldots,\chi_{i}$ could be represented as 
$$\tau^T_i(\chi_0,\ldots,\chi_i)=\chi_i+\varphi^T_i(\chi_0,\ldots,\chi_{i-1}),$$ 
where $\varphi^T_i$
is a Boolean function. 
The mapping $T$ is compatible  and ergodic iff,
additionally, the Boolean function
$\varphi^T_i$ is of odd weight, that is,
takes value $1$ exactly at the odd number of points 
$(\varepsilon_0,\dots,\varepsilon_{i-1})$, where
$\varepsilon_j\in\{0,1\}$ for $j=0,1,\ldots,i-1$. The latter holds if and only
if $\varphi^T_0=1$ and degree of 
$\varphi^T_i$ for
$i\ge 1$ is exactly
$i$, that is, the ANF of $\varphi^T_i$ contains a monomial
$\chi_0\cdots\chi_{i-1}$.
\end{thm}
\begin{cor}
\label{cor:Num}
There are exactly $2^{2^n-n-1}$
compatible and transitive 
mappings of $\mathbb Z/2^n$ onto $\mathbb Z/2^n$.
\end{cor}
From theorem \ref{ergBool} follows
an easy way to produce new ergodic functions out of given ones:
\begin{prop}
\label{compBool}
For any ergodic $f$ and any compatible $v$ the following functions
are ergodic: $f(x+4\cdot v(x))$, $f(x\oplus (4\cdot v(x)))$, $f(x)+4\cdot
v(x)$, and
$f(x)\oplus  (4\cdot v(x))$.
\end{prop}

With the use of theorem \ref{ergBool} one can determine whether a
given compatible mapping $f$ preserves measure (or is ergodic) assuming it
is bijective (respectively, transitive) modulo $2^n$ and studying behaviour
of the Boolean function $\delta_{n}(f)$. This approach 
is called a bit-slice analysis in \cite{KlSh}, \cite{KlSh-2}, and \cite{KlSh:3}.
More `analytic' techniques based on $p$-adic differential calculus and
Mahler interpolation series were developed in \cite{me-conf}, \cite{me-1},
and \cite{me-2};
see also \cite{Lar},\cite{Kot} and \cite{me-04} for various examples of compatible
and ergodic functions, 
for instance:
\begin{itemize}
\item (see \cite{me-conf}, \cite{me-1}) The function $f(x)=a+a_1(x\oplus
b_1)+\cdots+a_k(x\oplus b_k)$ is ergodic iff it is transitive modulo 4;
\item (see \cite{me-conf}, \cite{me-1}) The function $f(x)=a+a_0\cdot\delta_0(x)+a_1\cdot\delta_1(x)+\cdots$
is compatible and ergodic iff $a\equiv 1\pmod 2$, $a_0\equiv 1\pmod 4$,
and $a_i\equiv 0\pmod {2^i}$, $a_i\not\equiv 0\pmod {2^{i+1}}$ for $i=1,2,\ldots$;
\item (see \cite{Kot}) The function
$$f(x)=(\ldots((((x+c_0)\oplus d_0)+c_1)\oplus d_1)+\cdots +c_m)\oplus
d_m,$$  is ergodic iff $f$ is transitive modulo 4;
\item (see \cite{KlSh}) The function $f(x)=x+((x^2)\OR c)$ is ergodic iff $c\equiv 5\pmod 8$ or
$c\equiv 7\pmod 8$ (an equivalent statement --- iff $f$ is transitive modulo
8);
\item (see \cite{Lar}) The polynomial $f(x)=a_0 +a_1 x + \cdots+a_d x^d$ with integral
coefficients is ergodic iff the following congruences hold simultaneously:
\begin{gather*}
a_3+a_5+a_7+a_9+\cdots\equiv 2a_2\pmod 4;\
a_4+a_6+a_8+\cdots\equiv a_1+a_2-1\pmod 4;\\
a_1\equiv 1\pmod 2;\
a_0\equiv 1\pmod 2
\end{gather*}
(an equivalent statement --- iff $f$ is transitive modulo
8);
\item (see \cite{me-2}) A polynomial of degree $d$ with rational (and not
necessarily integral) coefficients 
is integer-valued,
compatible, and ergodic 
iff $f$ 
takes integral values at the points
$$0,1,\ldots,2^{\lfloor
\log_2 (\deg f)\rfloor +3}-1,$$ and the mapping
$$z\mapsto f(z)\bmod 2^{\lfloor
\log_2 (\deg f)\rfloor +3},$$ 
is compatible and transitive 
on the residue class ring  $\mathbb Z\big/
2^{\lfloor\log_2 d\rfloor +3}$ (i.e., 
modulo the biggest power
of 2 not exceeding $8 d$); 
\item (see \cite{me-conf}, \cite{me-1}) The entire function $f(x)=\frac{u(x)}{1+2\cdot
v(x)}$, where $u(x),v(x)$ are polynomials with integral coefficients, is
ergodic iff it is transitive modulo 8;
\item (see \cite[Example 3.6]{me-04}) The function $f(x)=ax+a^x$ is ergodic iff $a$ is odd
(an equivalent statement --- iff $f$ is transitive modulo 2).
\end{itemize}

A multivariate case was studied in \cite{KlSh:3},
\cite{me-04a}; see also \cite[Theorem 3.11]{me-2}. Multivariate ergodic
mappings could be of use in order to produce longer periods out of
shorter words operations: For instance, to obtain a period of length
$2^{256}$
one may use either univariate ergodic functions (hence,
$256$-bit operands)
or he may use $8$-variate ergodic functions and work with $32$-bit words.
Multivariate
ergodic mappings of \cite{KlSh:3} are conjugate to univariate ones (see \cite{me-04a});
hence
{\slshape despite all further results are stated for a univariate case,
they hold for these multivariate mappings
as well}. Thus 
a designer could use further constructions
either with longer words organized into $1$-dimensional arrays, or with
shorter
words organized into arrays of bigger dimensions.

\section{Constructions}
\label{sec:Constr}
In this section we introduce a method 
to construct counter dependent pseudorandom 
generators 
out of 
ergodic and measure-preserving mappings. The
method guarantees that
output sequences of these generators are always strictly uniformly
distributed. Actually, all these constructions are wreath
products of automata in the sense of \ref{def:WP}; the following results give us conditions these automata
should satisfy to produce a uniformly distributed output sequence. Our main technical tool is the following
\begin{thm}
\label{thm:WP}
Let $\mathcal G=\{g_0,\ldots,g_{m-1}\}$ be a finite sequence of 
compatible measure preserving
mappings of $\mathbb Z_2$ onto itself such that
\begin{enumerate}
\item the sequence $\{(g_{i\bmod m}(0))\bmod 2\colon i=0,1,2,\ldots\}$ is 
purely periodic, its shortest period is of length $m$;
\item $\sum_{i=0}^{m-1}g_i(0)\equiv 1\pmod 2$;
\item $\sum_{j=0}^{m-1}\sum_{z=0}^{2^k-1}g_j(z)\equiv 2^{k}\pmod {2^{k+1}}$
for all $k=1,2,\ldots$ .
\end{enumerate}
Then the recurrence sequence $\mathcal Z$ defined by the relation $x_{i+1}=g_{i\bmod
m}(x_i)$ is strictly uniformly distributed modulo $2^n$ for all $n=1,2,\ldots:$
That is, modulo each $2^n$ the sequence $\mathcal Z$  is purely periodic, its
shortest period is 
of
length 
$2^nm$, and each element of $\mathbb Z/2^n$ occurs at the period
exactly $m$ times. 
\end{thm}
\begin{note*}
In view of \ref{ergBool}
condition  (3)
of theorem \ref{thm:WP} could be replaced by the equivalent condition
$$\sum_{j=0}^{m-1}\Coef_{0,\ldots,k-1}(\varphi_k^j)\equiv 1\pmod 2 \qquad (k=1,2,\ldots),$$
where $\Coef_{0,\ldots,k-1}(\varphi)$
is a coefficient of the monomial $\chi_0\cdots\chi_{k-1}$ in the
Boolean polynomial $\varphi$.
\end{note*}
It turns out that the sequence $\mathcal Z$ of \ref{thm:WP} is just the sequence $\mathcal
Y$ of the following
\begin{lem}
\label{le:WP-odd} 
Let $c_0,\ldots,c_{m-1}$ be a finite sequence of $2$-adic integers, and
let $g_0,\ldots,g_{m-1}$ be a finite sequence of compatible
mappings of $\mathbb Z_2$ onto itself such that 
\renewcommand{\theenumi}{\roman{enumi}}
\begin{enumerate}
\item $g_j(x)\equiv x+c_j\pmod 2$ for $j=0,1,\ldots,m-1$, 
\item $\sum_{j=0}^{m-1}c_j\equiv
1\pmod 2$, 
\item the sequence 
$\{c_{i\bmod m}\bmod 2\colon i=0,1,2,\ldots\}$ is purely periodic, its shortest period is
of length $m$,
\item $\delta_k(g_j(z))\equiv \zeta_k+\varphi_k^j(\zeta_0,\ldots,\zeta_{k-1})\pmod
2$, $k=1,2,\ldots$,
where $\zeta_r=\delta_r(z)$, $r=0,1,2,\ldots$, 
\item for each $k=1,2,\ldots$ an odd number of Boolean polynomials 
$\varphi_k^j$
in the Boolean variables $\zeta_0,\ldots,\zeta_{k-1}$ are of odd weight.
\end{enumerate}
\renewcommand{\theenumi}{\arabic{enumi}}
Then the recurrence sequence $\mathcal Y=\{x_i\in\mathbb Z_2\}$ defined by the relation 
$x_{i+1}=g_{i\bmod m}(x_i)$ is strictly uniformly distributed: 
It is purely periodic modulo $2^k$ for all $k=1,2,\ldots$; its  shortest period is
of length $2^km$; each element of $\mathbb Z/2^k$ occurs at
the period exactly $m$ times. 
Moreover,
\begin{enumerate}
\item  
the sequence 
$\mathcal D_s=\{\delta_s(x_i)\colon i=0,1,2,\ldots\}$
is purely periodic; it has a period of length $2^{s+1}m$,
\item $\delta_s(x_{i+2^{s}m})\equiv\delta_s(x_{i})+1\pmod
2$ for all $s=0,1,\ldots, k-1$, $i=0,1,2,\ldots$, 
\item for each $t=1,2,\ldots,k$ and each $r=0,1,2,\ldots$ the sequence 
$$x_r\bmod 2^t,x_{r+m}\bmod 2^t,x_{r+2m}\bmod 2^t,\ldots$$
is purely periodic, its shortest period is of length $2^t$, each element
of $\mathbb Z/2^t$ occurs at the period exactly once.
\end{enumerate} 
\end{lem}
%
\begin{note}
\label{note:Bool} 
Assuming $m=1$ in \ref{thm:WP} one obtains ergodicity criterion
\ref{ergBool}.
\end{note}
\begin{cor}
\label{cor:WP}
Let  a finite sequence  of mappings $\{g_0,\ldots,g_{m-1}\}$ of $\mathbb
Z_2$ into itself satisfy conditions
of theorem \ref{thm:WP}, and let $\{F_0,\ldots,F_{m-1}\}$ be an arbitrary
finite sequence of balanced {\rm (}and not necessarily compatible{\rm
)} mappings of $\mathbb Z/2^n$ $(n\ge 1)$ onto $\mathbb
Z/2^k$, $1\le k\le n$. Then the sequence
$\mathcal F=\{F_{i\bmod m}(x_i)\colon i=0,1,2\ldots\}$, where $x_{i+1}=g_{i\bmod m}(x_i)\bmod
2^n$, is strictly uniformly distributed over $\mathbb Z/2^k:$ It is purely
periodic with a period of length $2^nm$, and each element of $\mathbb Z/2^k$
occurs at the period  exactly $2^{n-k}m$ times.
\end{cor}

Theorem \ref{thm:WP} and lemma \ref{le:WP-odd} together with corollary \ref{cor:WP} 
enables one to construct a counter-dependent generator out of the following components:
\begin{itemize}
\item A sequence $c_0,\ldots,c_{m-1}$ of integers, which we call a \textit{control
sequence}.
\item A sequence $h_0,\ldots,h_{m-1}$ of compatible mappings, which is
used to form a sequence of clock state update functions $g_i$ (see e.g.
examples \ref{WP-even}).
\item A sequence $H_0,\ldots,H_{m-1}$ of compatible mappings to produce
clock output functions $F_i$ (see e.g. proposition \ref{prop:reverse}).
\end{itemize}
Note that ergodic functions that are needed to meet conditions of \ref{prop:reverse}
or \ref{WP-even} (3) could be produced out of compatible ones with the use of \ref{Delta}
or \ref{compBool}.
A control sequence could be produced by an external generator (which
in turn could be a generator of the kind considered in this paper), or it
could be just a queue the state update and output functions are called
from a look-up table.
The functions $h_i$ and/or $H_i$  could be either precomputed to arrange
that look-up table, 
or they could be produced on-the-fly in a form that is determined
by a control sequence. This form may also look  `crazy', e.g., 
\begin{equation}
\label{eq:crazy}
h_i(x)=(\cdots((u_0(\delta_0(c_i))\bigcirc _{\delta_1(c_i),
\delta_2(c_i)}u_1(\delta_3(c_i)))\bigcirc _{\delta_4(c_i),\delta_5(c_i)}u_2(\delta_6(c_i)))\cdots,
\end{equation}
where $u_j(0)=x$, the variable, and $u_j(1)$ is a constant (which is determined
by $c_i$, or is read from a precomputed look-up table, etc.), while (say) $\bigcirc _{0,0}=+$,
an integer addition, $\bigcirc _{1,0}=\cdot$, an integer multiplication,
$\bigcirc _{0,1}=\XOR$, $\bigcirc _{1,1}=\AND$.  There is absolutely no matter
what these $h_i$ and $H_i$ look like or how they are obtained, 
{\slshape the above stated results give a general method to combine all the data together
to produce a uniformly distributed output sequence of a maximum period
length}. 
\begin{exmp}
\label{WP-even}
These are obtained with the use of \ref{le:WP-odd}, \ref{ergBool}, \ref{compBool},
and \eqref{eq:sumBool}.

\begin{enumerate}
\item A control sequence could be produced by the generator ${\mathfrak A}=\langle \mathbb
Z/2^s, \mathbb Z/2^s,f,F,u_0\rangle$ (see Section \ref{Prelm}) with ergodic
state update function $f$ and measure-preserving
output function $F$. Then length of the shortest period of the control sequence is $m=2^s$,
see \ref{prop:Auto}. Take $m$ arbitrary ergodic functions $h_0,\ldots,h_{m-1}$ and arbitrary
odd $k\in\{0,1,\ldots,m-1\}$, and 
put
$\breve g_0(x)=x\oplus(x+1)\oplus h_0(x),\ldots,\breve g_{k-1}=x\oplus(x+1)\oplus h_{k-1}(x)$, 
$\breve g_k=h_k,\ldots,\breve g_{m-1}=h_{m-1}$, $g_i=\breve g_{c_i\bmod
m}$ for $i=0,1,2,\ldots$. In other words, in this case the control sequence
just define the queue the functions $\breve g_j$ are called,
thus producing the output sequence 
$$x_0,x_1=\breve g_{c_0}(x_0)\bmod 2^n, x_2=\breve g_{c_1}(x_1)\bmod2^n,
\ldots$$
Obviously, in this example a control sequence could be an arbitrary permutation
of $0,1,\ldots,2^s-1$, and not necessarily an output of the generator $\mathfrak
A$.
\item
Now let 
$\{c_0,\ldots,c_{m-1}\}$ be an arbitrary 
sequence
of length $m=2^s$, i.e.,  $c_0,\ldots,c_{m-1}$ are not necessarily pairwise
distinct.  
Let $\{h_0,\ldots,h_{m-1}\}$ be arbitrary compatible and ergodic
mappings.
For $0\le j\le m-1$ put $g_j(x)=c_j+
h_j(x)$. 
\footnote{one may also put $g_j(x)=(c_j+x)\oplus(2\cdot h_j(x))$.}
These mappings
$g_j$ satisfy
conditions of theorem \ref{thm:WP}
if and only if
$\sum_{j=0}^{2^m-1}c_j\equiv 1\pmod 2$.
\newpage 
\item
For $m>1$ odd let $\{h_0,\ldots,h_{m-1}\}$ be a finite sequence of compatible 
and ergodic mappings; 
let $\{c_0,\ldots,c_{m-1}\}$  be a finite sequence of 
integers such that 
\begin{itemize}
\item $\sum_{j=0}^{m-1}c_j\equiv
0\pmod 2$, and 
\item the sequence 
$\{c_{i\bmod m}\bmod 2\colon i=0,1,2,\ldots\}$ is purely periodic 
with the shortest period of length $m$.
\end{itemize}
Put $g_j(x)=c_j\oplus h_j(x)$ {\rm (}respectively, $g_j(x)=c_j+h_j(x)${\rm)}.
Then $g_j$ satisfy conditions of \ref{thm:WP}.
\item The conditions of (3) are satisfied in the case $m=2^s-1$ and $\{c_0,\ldots,c_{m-1}\}$
is the output sequence 
of a maximum period
linear feedback
shift register over $\mathbb Z/2$ with $s$ cells.
\end{enumerate}
\end{exmp}

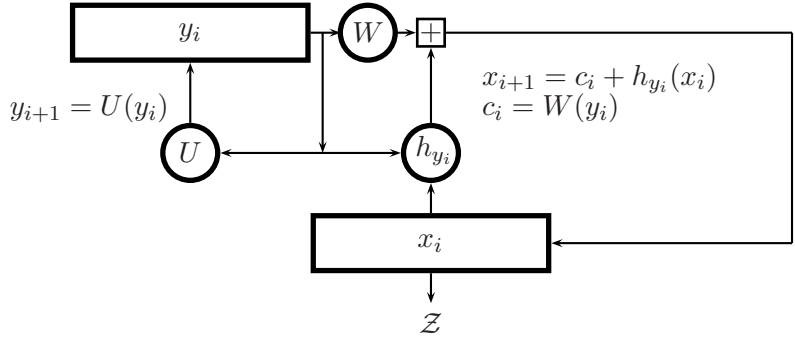
\begin{figure}
\begin{quote}\psset{unit=0.4cm}
 \begin{pspicture}(-6,5)(24,15)
\psframe[linewidth=1pt](11.5,13.5)(12.5,14.5)
\pscircle[linewidth=2pt](12,10){1}
\pscircle[linewidth=2pt](4,10){1}
\pscircle[linewidth=2pt](9.9,14){1}
\psline{->}(8.4,14)(8.4,10)
\psline{<->}(11,10)(5,10)
\psline{->}(4,11)(4,13)
  \psline{->}(12,11)(12,13.5)
  \psline{->}(24,7)(16,7)
  \psline{->}(8,14)(9,14)
  \psline{->}(10.8,14)(11.5,14)
  \psline(24,14)(24,7)
  \psline(12.5,14)(24,14)
  \psline{->}(12,8)(12,9)
  \psline{<-}(12,5)(12,6)
  \psframe[linewidth=2pt](0,13)(8,15)
  \psframe[linewidth=2pt](8,6)(16,8)
  \uput{0}[180](12.4,7){$x_i$}
  \uput{0}[90](12.1,9.6){$h_{y_i}$}
  \uput{0}[90](9.9,13.7){$W$}
  \uput{0}[90](4,9.7){$U$}
  \uput{0}[90](12,13.7){+}
  \uput{1}[90](12,3){$\mathcal Z$}
  \uput{1}[0](12.7,12.5){
  $x_{i+1}
  =c_i+h_{y_i}(x_i)$}
  \uput{1}[0](12.7,11.5){
  $c_i
  =W(y_i)
  $}
  \uput{1}[0](-3,11.5){
  $y_{i+1}=U(y_i)$}
  \uput{1}[0](2.6,14){$y_i$}
 \end{pspicture}
\end{quote}
\caption{Wreath product basic circuit of Examples \ref{WP-even}, (2)--(4).} 
\label{fig:wr}
\end{figure}

A basic circle illustrating these example wreath products is given at Figure
\ref{fig:wr}.
A number of counter dependent generators could be derived from \ref{WP-even}
by taking explicit expressions for involved mappings. 
For instance, one can obtain the following result,
which is a variation of theme of \cite[Theorem 3]{KlSh-2}).
Take odd $m>1$ and consider  a finite sequence $C_0,\ldots,C_{m-1}$
of integers
such that $\delta_0(C_j)=1$ and $\delta_2(C_j)=1$, $j=0,1,\ldots, m-1$. Let
a sequence $\{c_j\colon j=0,1,2,\ldots\}$ satisfy conditions of \ref{WP-even}(3).  
Then the sequence $\{x_{i+1}=(x_i+c_i+(x_i^2\OR C_i))\bmod 2^n\colon i=0,1,2,\ldots\}$
is purely periodic modulo $2^k$ for all $k=1,2,\ldots$
with the shortest period of length $2^km$, and each element of $\mathbb Z/2^k$ occurs at
the period exactly $m$ times.
This is a stronger claim in comparison with that of \cite[Theorem 3]{KlSh-2}): 
Not only the sequence of pairs $(y_i,
x_i)$ defined by $y_{i+1}=(y_i+1)\bmod m$; $x_{i+1}=(x_i+c_i+(x_i^2\OR C_{y_i}))\bmod 2^n$
is periodic with a period of length $2^nm$, yet length of the shortest period of the sequence
$\{x_i\}$ is $2^nm$. The latter could never be achieved under conditions
of Theorem 3 of \cite{KlSh-2}: They imply that the length of the shortest period of the
sequence $\{x_i\pmod 2\}$ is $2$, and not $2m$.

%
\section{Properties of output sequences}
\label{Out}

\subsection*{Distribution of $k$-tuples} 
The output sequence $\mathcal Z$ of any wreath product
of automata that satisfy 
\ref{thm:WP} 
is strictly uniformly
distributed as a sequence over $\Z/2^n$ for all $n$. That is, each sequence
$\mathcal Z_n$ of residues modulo $2^n$ of members of the sequence $\mathcal Z$
is purely periodic,
and each element of $\Z/2^n$
occurs at the period the same number of times. However, when this sequence
$\mathcal Z_n$
is used as a key-stream, that is, 
as a binary sequence $\mathcal Z^\prime_n$ obtained by a concatenation of successive $n$-bit words
of $\mathcal Z$, it
is important to know how
$n$-tuples are distributed in this binary sequence. 
Yet strict uniform distribution of an
arbitrary sequence $\mathcal T$
as a sequence over $\Z/2^n$ does not necessarily imply uniform distribution
of $n$-tuples, if this sequence is considered as a binary sequence
$\mathcal T^\prime$.

For instance, let $\mathcal T=023102310231\ldots$. This sequence is strictly uniformly
distributed over $\Z/4$; the length of its shortest period is $4$. Its binary representation
is $\mathcal T^\prime_2=000111100001111000011110\ldots$ Considering $\mathcal T$ as
a sequence over $\Z/4$, each number of $\{0,1,2,3\}$ occurs in the
sequence with the same frequency $\frac{1}{4}$. Yet if we consider $\mathcal
T$ in its binary form $\mathcal T^\prime_2$, then $00$ (as well as $11$) occurs in this sequence with
frequency $\frac{3}{8}$, whereas $01$ (as well as $10$) occurs with frequency $\frac{1}{8}$.

In this subsection we show that such an effect does not take
place for output sequences of automata described in 
\ref{thm:WP}, \ref{le:WP-odd},
and \ref{WP-even}: 
{\slshape Considering any of these sequences in a
binary form, a distribution
of $k$-tuples is uniform, for all $k\le n$}. Now we state this property  formally.

%
Consider a (binary) {\it $n$-cycle} 
$C=(\varepsilon_0\varepsilon_1\dots \varepsilon_{n-1})$, i.e., 
an oriented
graph on vertices $\{a_0,a_1,\ldots, a_{n-1}\}$ and edges 
$$\{(a_0,a_1),(a_1,a_2),\ldots, (a_{n-2},a_{n-1}),(a_{n-1},a_0)\},$$ 
where
each vertex $a_j$ is labelled with $\varepsilon_j\in\{0,1\}$, $j=0,1,\dots,n-1$.
(Note that then $(\varepsilon_0\varepsilon_1\dots \varepsilon_{n-1})=
(\varepsilon_{n-1}\varepsilon_0\dots \varepsilon_{n-2})=\ldots$, etc.).
Clearly, each purely periodic sequence $\mathcal S$ over $\Z/2$ with period 
$\alpha_0\ldots\alpha_{n-1}$
of length $n$
could be related to a binary $n$-cycle $C(\mathcal S)=(\alpha_0\ldots\alpha_{n-1})$.
Conversely, to each binary $n$-cycle $(\alpha_0\ldots\alpha_{n-1})$ we could
relate $n$ purely periodic binary sequences with periods of length $n$: Those
are $n$ shifted versions of the sequence
$$\alpha_0\ldots\alpha_{n-1}\alpha_0\ldots\alpha_{n-1}\ldots.$$

Further, {\it a $k$-chain in a binary $n$-cycle}  
$C$ is a
binary string $\beta_0\dots\beta_{k-1}$, $k<n$, that satisfies the following
condition: There exists $j\in\{0,1,\ldots,n-1\}$ such that $\beta_i=\varepsilon_{(i+j)\bmod
n}$ for $i=0,1,\ldots, k-1$. Thus, a $k$-chain
is just a string of length
$k$ of labels that corresponds to a chain of length $k$ in a graph $C$.
We call a binary $n$-cycle $C$ {\it $k$-full}, if each $k$-chain
occurs in the graph $C$ the same number $r>0$ of times.

Clearly, if $C$ is $k$-full, then $n=2^kr$. For instance, a well-known
De Bruijn sequence is an $n$-full $2^n$-cycle. 
Clearly enough that a $k$-full $n$-cycle is $(k-1)$-full:
Each $(k-1)$-chain occurs in $C$ exactly $2r$ times, etc. Thus, if an $n$-cycle
$C(\mathcal S)$ is $k$-full, then each $m$-tuple (where $1\le m\le k$) occurs in
the sequence $\mathcal S$ with the same probability (limit frequency) $\frac{1}{2^m}$.
That is, the sequence $\mathcal S$ is {\it $k$-distributed}, see
\cite[Section 3.5, Definition D]{Knuth}.
\begin{defn} A purely periodic binary sequence $\mathcal S$ with the shortest
period of length
$N$ is said to be {\it
strictly $k$-distributed} iff the corresponding $N$-cycle $C(\mathcal S)$
is $k$-full.
\end{defn}

Thus, if a sequence $\mathcal S$ is strictly $k$-distributed, then it is
strictly $s$-distributed, for all positive $s\le k$.

\begin{thm} For the sequence $\mathcal Z$ of theorem \ref{thm:WP} each binary sequence
$\mathcal Z^\prime_n$ is
strictly $k$-distributed
for all $k=1,2,\ldots,n$. 
\label{thm:distr}
\end{thm} 

\begin{note}
\label{note:distr} 
Theorem \ref{thm:distr} remains true 
for the sequence $\mathcal F$ of corollary
\ref{cor:WP}, where $F_j(x)=\big\lfloor\frac{x}{2^{n-k}}\big\rfloor\bmod 2^k$, 
$j=0,1,\ldots,m-1$, a truncation of $(n-k)$ less significant bits. Namely, {\it
a binary representation $\mathcal F^\prime_n$ of the sequence $\mathcal F$
is a purely periodic strictly $k$-distributed binary sequence with a period
of length $2^nm k$.}
\end{note}
Theorem \ref{thm:distr} treats an output sequence of a counter-dependent
automaton as an infinite (though, a periodic) binary sequence. However, in cryptography
only a part of a period is used during encryption. So it is natural
to ask how `random' is a finite segment (namely, the period) of this infinite sequence.
According to \cite[Section 3.5, Definition Q1]{Knuth} 
a finite binary sequence 
$\varepsilon_0\varepsilon_1\dots \varepsilon_{N-1}$
of length $N$ is said to be random, iff
\begin{equation}
\label{eq:Q1}
\bigg|\frac{\nu(\beta_0\ldots\beta_{k-1})}{N}-\frac{1}{2^k}\bigg|\le\frac{1}{\sqrt
N}
\end{equation}
for all $0<k\le\log_2N$, where $\nu(\beta_0\ldots\beta_{k-1})$ is the number
of occurrences of a binary word $\beta_0\ldots\beta_{k-1}$ in a binary word
$\varepsilon_0\varepsilon_1\dots \varepsilon_{N-1}$. If a finite sequence
is random in the sense of this Definition Q1 of \cite{Knuth}, we shall say
that this sequence 
{\it satisfies} Q1. We shall also
say that an {\it infinite periodic sequence satisfy} Q1 iff its shortest
period satisfies Q1.
Note that, contrasting to the case of strict $k$-distribution, which implies
strict $(k-1)$-distribution, 
it is not enough to demonstrate only
that \eqref{eq:Q1}
holds for $k=\lfloor\log_2N\rfloor$ to prove a finite sequence of length $N$ 
satisfies Q1:
For instance, the sequence $1111111100000111$ satisfies \eqref{eq:Q1} for
$k=\lfloor\log_2N\rfloor=4$ and does not satisfy \eqref{eq:Q1} for $k=3$.

\begin{cor} 
\label{cor:distr}
The sequence $\mathcal
Z^\prime_n$ of theorem \ref{thm:distr} 
satisfies {\rm Q1} if $m\le\frac{2^n}{n}$. Moreover, in this case 
under the conditions of \ref{note:distr} the output binary sequence still
satisfies {\rm Q1}
if one truncates 
$0\le k\le\frac{n}{2}-\log_2\frac{n}{2}$ 
lower order bits {\rm (that is, if one uses clock output functions $F_j$
of \ref{note:distr})}.
\end{cor}
We note here that according to \ref{cor:distr} a control sequence
of a counter-dependent automaton (see \ref{thm:WP}, \ref{le:WP-odd}, \ref{cor:WP},
and the text and examples thereafter) may not satisfy Q1 at all, yet nevertheless
a corresponding output sequence necessarily satisfies Q1. Thus, {\slshape
with the use of
wreath product
techniques one could stretch `non-randomly looking' sequences to `randomly
looking' ones}.

\subsection*{Structure}
A recurrence sequence could be `very uniformly distributed', yet nevertheless could
have some mathematical structure that might be used by
an attacker to break the cipher. For instance, a clock sequence
$x_i=i$ is uniformly distributed in $\mathbb Z_2$; moreover,
its counterpart in the field $\mathbb R$ of real numbers, the so-called Van der Corput
sequence $u_i={i}\cdot{2^{-\lfloor\log_2i\rfloor-1}}$,
has the least (of the known)
discrepancy, see \cite{KN}. We are going to study what structure could
have sequences outputted by our counter-dependent generators. 

Theorem \ref{thm:WP} immediately implies that the
{\it $j$\textsuperscript{th} coordinate sequence $\delta_j(\mathcal Z)=\{\delta_j(x_i)\colon i=0,1,2,\ldots\}$
$(j=0,1,2,\ldots)$ of the sequence $\mathcal Z$}, i.e., a sequence
formed by all $j$\textsuperscript{th} bits of members
of the sequence $\mathcal Z$,  
has a period not longer than $m\cdot 2^{j+1}$. Moreover, the following 
could be easily proved: 
\begin{prop}
\label{note:halfper-odd}
{\rm (1)} The $j$\textsuperscript{th} coordinate sequence
$\delta_j(\mathcal Z)$
is a purely periodic binary sequence with a period of length $2^{j+1}m$, and {\rm
(2)} the
second half of the period is a bitwise negation of the first half: $\delta_j(x_{i+2^jm})\equiv
\delta_j(x_i)+1\pmod 2$, $i=0,1,2,\ldots$  
\end{prop}
This means
that
the $j$\textsuperscript{th} coordinate sequence of the sequence of states of
a counter-dependent generator is completely determined by
the first half of its period; so, intuitively, it is as
`complex' as the first half of its period. Thus we ought to understand what
sequences of length $2^jm$ occur as the first half of the
period of the $j$\textsuperscript{th} coordinate sequence. 

For $j=0$ (and
$m>1$) the
answer immediately follows from \ref{thm:WP} and \ref{le:WP-odd} --- any binary
sequence $c_0,\ldots,c_{m-1}$ such that $\sum_{j=0}^{m-1}c_j\equiv
1\pmod 2$ does.  It turns out that for $j>0$ 
{\slshape any binary sequence could be produced as
the first half of the period of the $j$\textsuperscript{th} coordinate sequence independently
of other coordinate sequences}.

More formally, 
to each sequence $\mathcal Z$ described by theorem \ref{thm:WP}
we associate a sequence $\Gamma(\mathcal Z)=\{\gamma_1,\gamma_2,\ldots\}$
of non-negative rational integers $\gamma_j$ such that $0\le\gamma_j\le
2^{2^{j}m}-1$ 
and the base-$2$ expansion of $\gamma_j$
agrees with the first half
of the period of the $j$\textsuperscript{th} coordinate sequence $\delta_j(\mathcal
Z)$ for all $j=1,2,\ldots$; that is
$$\gamma_j=\delta_j(x_0)+2\cdot\delta_j(x_1)+
4\cdot\delta_j(x_2)+\dots+2^{2^jm-1}\cdot\delta_j(x_{2^jm-1}),$$
where $x_0$ is an initial state; $x_{i+1}=g_{i\bmod m}(x_i)$, $i=0,1,2,\ldots$.
Now
we take an arbitrary sequence 
$\Gamma(\mathcal Z)=\{\gamma_1,\gamma_2,\ldots\}$
of non-negative rational integers $\gamma_j$ such that $0\le\gamma_j\le
2^{2^{j}m}-1$ 
and wonder
whether this sequence could be so associated to some sequence $\mathcal Z$
described by theorem \ref{thm:WP}.

%
%
The answer is {\slshape yes}. Namely, the
following theorem holds.
\begin{thm}
\label{thm:WP:AnyHalfPer}
Let $m> 1$ be a rational integer, and let $\Gamma=\{\gamma_1,\gamma_2,\dots\}$ 
be an arbitrary sequence over $\N_0$
such that $\gamma_j\in\{1,2,\ldots,2^{2^jm}-1\}$ for
all $j=1,2,\dots$. Then there exist  a finite sequence 
$\mathcal G=\{g_0,\ldots,g_{m-1}\}$
of compatible  measure preserving mappings of $\Z_2$ onto itself and a
$2$-adic integer $x_0=z\in\Z_2$ such that $\mathcal G$ satisfies conditions
of theorem \ref{thm:WP}, and the base-$2$ expansion of $\gamma_j$ agrees
with the first $2^jm$ terms of the sequence $\delta_j(\mathcal Z)$ 
for all $j=1,2,\dots$,
where the recurrence sequence $\mathcal Z=\{x_0,x_1,\ldots\in\Z_2\}$ is
defined by the recurrence relation $x_{i+1}=g_{i\bmod m}(x_i)$, $(i=0,1,2,\dots)$.
In the case $m=1$ the assertion holds for an arbitrary $\Gamma=\{\gamma_0,\gamma_1,\dots\}$,
where $\gamma_j\in\{1,2,\ldots,2^{2^j}-1\}$, $j=0,1,2,\dots$.
\end{thm} 
\subsection*{Linear complexity} The latter is an important cryptographic
measure of
complexity of a binary sequence; being a number of cells of the
shortest linear feedback shift register (LFSR) that outputs the given sequence%
\footnote{i.e., degree of the minimal polynomial over $Z/2$ of the given
sequence} 
it estimates dimensions of a linear system an attacker must solve to
obtain initial state. 
\begin{thm}
\label{thm:lincomp:sharp} For $\mathcal Z$ and $m$ of theorem \ref{thm:WP} let
$\mathcal Z_j=\delta_j(\mathcal Z)$, $j>0$, be
the $j$\textsuperscript{th}
coordinate sequence. 
Represent $m=2^kr$, where $r$
is odd.
Then length of the shortest period of $\mathcal Z_j$ is $2^{k+j+1}s$ for 
some $s\in\{1,2,\dots,r\}$, and both extreme cases $s=1$ and $s=r$ 
occur: For every sequence $s_1,s_2,\ldots$ over a set $\{1,r\}$
there exists a sequence $\mathcal Z$ of theorem \ref{thm:WP}
such that length of the shortest period  of
$\mathcal Z_j$ is 
$2^{k+j+1}s_j$, $(j=1,2,\ldots)$.
Moreover, linear complexity $\Psi_2(\mathcal Z_j)$ 
of the sequence $\mathcal
Z_j$ 
satisfies the following inequality:
$$2^{k+j}+1\le \Psi_2(\mathcal Z_j)\le 2^{k+j}r+1.$$
Both these bounds are sharp: 
For every sequence $t_1,t_2,\ldots$ over a set $\{1,r\}$ 
there exists a sequence $\mathcal Z$ of theorem \ref{thm:WP}
such that linear complexity of
$\mathcal Z_j$ is exactly
$ 2^{k+j}t_j+1$, $(j=1,2,\ldots)$.
\end{thm}
\begin{note*} Somewhat similar estimates hold for $2$-adic span (see definition
in \cite{Kl-Gor}),
one more cryptographic
measure of complexity of a sequence. We have to omit exact statements due
to space limitations.
\end{note*}
Whereas the linear complexity of a binary sequence $\mathcal X$ is the length of the shortest LFSR 
that produces $\mathcal X$,
the {\it $\ell$-error linear complexity} is the length of the  shortest LFSR  that produces a sequence with almost the same (with the exception of not more than $\ell$ members)  period  as that of $\mathcal
X$; that is, the two periods coincide everywhere  but at $t\le\ell$
places.
Obviously, a random sequence of length $L$
coincides with a sequence that has a period of length $L$ approximately at
$\frac{L}{2}$ places. That is, the $\ell$-error linear complexity makes sense
only for $\ell<\frac{L}{2}$. The following proposition holds.
\begin{prop}
\label{prop:coord}
Let $\mathcal Z$ 
be a sequence of Theorem \ref{thm:WP}, and let $m=2^s>1$. 
Then for $\ell$  less than the half of the length of the shortest
period of the $j$-th coordinate sequence $\delta_j(\mathcal Z)
$,
the $\ell$-error linear complexity of $\delta_j(\mathcal Z)$ exceeds $2^{j+m-1}$, the half of the length of its
shortest period.
\end{prop}


From \ref{thm:lincomp:sharp} it follows that the less is $j$, the shorter is a period (and the smaller is 
linear
complexity) of the
coordinate sequence $\mathcal Z_j$. 
This could be improved by truncation of less significant bits (see \ref{cor:distr}) or, if necessary,
with the use of clock output functions of special kind:
\begin{prop}
\label{prop:reverse}
Let $H_i\colon\Z_2\rightarrow\Z_2$ $(i=0,1,2,\ldots,m-1)$
be compatible and ergodic mappings. 
For $x\in\{0,1,\ldots,2^n-1\}$ let $F_i(x)=(H_i(\pi(x)))\bmod 2^n$,
where $\pi$ is a permutation of bits of  $x\in\Z/2^n$ such that $\delta_0(\pi(x))=\delta_{n-1}(x)$. 
Consider a sequence $\mathcal F$ of \ref{cor:WP}.
Then  the shortest period of the $j$\textsuperscript{th}
coordinate sequence $\mathcal F_j=\delta_j(\mathcal F)$ $(j=0,1,2,\dots,n-1)$ 
is of length $2^nk_j$ for a suitable $1\le k_j\le m$. Moreover, linear complexity of
the sequence $\mathcal F_j$ exceeds $2^{n-1}$.
\begin{note*} In view of Note \ref{note:Bool}, all the results of Section
\ref{Out} remain true for compatible mappings $T\colon\Z_2\>\Z_2$ (i.e.,
for T-functions) either.
\end{note*}
%
\end{prop}

\section{Security issues}
\label{Sec}
The paper introduces design techniques that guarantees in advance that the
so constructed generator, which dynamically modifies itself during encryption, will
meet certain important cryptographic properties;
namely, long period, uniform distribution and high linear complexity of
the output sequence.
The techniques can not guarantee per se that every such cipher will be
secure --- obvious degenerative cases exist. 
On the other
hand, if clock state update functions $g_i$ are chosen arbitrarily
under the conditions of \ref{thm:WP}, and clock output functions $F_i$ just
truncate $k$ low order bits, $k\approx \frac{n}{2}$ (see \ref{cor:distr}),
theorem \ref{thm:WP:AnyHalfPer} leaves no chance to an attacker to break such
a scheme. Yet in practice we can not choose $g_i$ arbitrarily; restrictions
are determined by concrete implementations, which are not discussed here.
 
In this section we are going to give some evidence
that with the use of the techniques described above
it might be possible to design stream ciphers such that the problem of their
key recovery
is intractable up to the following conjecture:
Choose
(randomly and independently) $k\le n$  ANF's 
$\psi_i$
in $n$ Boolean variables $\chi_0,\ldots,\chi_{n-1}$ from the class of ANF's
with polynomially restricted number of monomials.
Consider
a mapping $F\:\Z/2^n\>\Z/2^k$:
$$F(x)=F(\chi_0,\ldots,\chi_{n-1})=\psi_0(\chi_0,\ldots,\chi_{n-1})\oplus
\psi_1(\chi_0,\ldots,\chi_{n-1})\cdot 2\oplus\dots
\oplus\psi_{k-1}(\chi_0,\ldots,\chi_{n-1})\cdot 2^{k-1},$$
where $\chi_j=\delta_j(x)$ for $x\in\Z/2^n$. We conjecture that {\slshape 
this function $F$ is one-way}, that is, 
one could invert it (i.e., could find an $F$-preimage in case it exists)
only with a negligible 
in $n$ 
probability. Note that to
find any $F$-preimage, i.e., to solve an equation $F(x)=y$ in unknown $x$
one has to solve a system of $k$ Boolean equations in $n$ variables.
Yet {\slshape to determine whether $k$ ANF 
have common zero is an $NP$-complete problem}, 
see e.g. \cite[Appendix
A, Section A7.2, Problem ANT-9]{GJ}.

Of course, it is not sufficient to conjecture $F$ is one-way in case we
only know that the problem
of whether $F$-preimage exists is $NP$-complete; 
it must be hard in average to invert $F$. However, to our best
knowledge, no polynomial-time algorithms that solve random systems of $k$
Boolean equations in $n$ variables for so restricted $k$ are known. The best known results are
polynomial-time
algorithms that solve so-called overdefined Boolean systems of degree
not more than 2, i.e., systems where the number
of equations is greater than the number of unknowns and where each ANF
is at most quadratic, see \cite{BFS}, \cite{XL}.

Proceeding with the above plausible conjecture, 
to each ANF 
$\psi_i$, $i=0,1,2,\ldots,k-1$ 
we relate a mapping $\Psi_i\:\Z_2\>\Z_2$
in the following way: $\Psi_i(x)=\psi_i(\delta_0(x),\ldots,\delta_{n-1}(x))\in\{0,1\}
\subset\Z_2$. 
Now to each above mapping $F$ we relate a mapping 
$$f_F(x)=(1+x)\oplus 2^{n+1}\cdot F(x)=(1+x)\oplus 2^{n+1}\cdot\Psi_0(x)\oplus 2^{n+2}\cdot\Psi_1(x)\oplus\dots\oplus 2^{n+k}\cdot\Psi_{k-1}(x)$$
of $\Z_2$ onto itself. 
%
%
%
Clearly,
$$
\delta_j(f_F(x))=\begin{cases}
1\oplus\delta_0(x),\qquad\text{if $j=0$;}\\
\delta_j(x)\oplus\delta_0(x)\cdots\delta_{j-1}(x),\qquad\text{if $0<j\le n$;}\\
\delta_j(x)\oplus\delta_0(x)\cdots\delta_{j-1}(x)\oplus
\psi_{j-n-1}(\delta_0(x),\dots,\delta_{n-1}(x)),\qquad\text{if
$n+1\le j\le n+k$.}
\end{cases}$$
In view of \ref{ergBool} the mapping $f_F\:\Z_2\>\Z_2$ is compatible and
ergodic for any choice of ANF's
$\psi_0,\ldots,\psi_{k-1}$.

Now for $m=2^n$ and $i=0,1,2,\ldots,m-1$ choose arbitrarily and independently mappings
$F_i\:\Z/2^n\>\Z/2^k$ of the above kind.
Put $d_0=\ldots=d_{2^n-3}=0$, $d_{2^n-2}=d_{2^n-1}=1$, and consider a recurrence
sequence of states
$x_{i+1}=d_{i\bmod m}\oplus f_{F_{i\bmod m}}(x_i)$ and a corresponding output
sequence $g(x_0),g(x_1),\ldots$ over $\Z/2^k$,
where 
$g(x)=\lfloor\frac{x}{2^{n+1}}\rfloor\bmod2^k$, a truncation. In view of
\ref{WP-even} the output sequence satisfy \ref{cor:WP}.
%

We shall always take a key $z\in\{0,1,\ldots,2^n-1\}$
as an initial state $x_0$. 
Let $z$ be the only information that is not known to an
attacker, let  everything else, i.e., $n$, $k$, $f_{F_i}$, $d_i$, and $g$, 
as well as  the first $s$ members of the output sequence $\{y_i\}$,
be known to him.
Since $\delta_0(x)\cdots\delta_{j-1}(x)=1$ iff $x\equiv -1\pmod{2^j}$,  
with probability  $1-\epsilon$
(where  $\epsilon$ is negligible if $s$ is a polynomial in $n$) he obtains
a sequence\footnote{which is pseudorandom even if $F=F_0=F_1=\ldots$, under additional conjecture (how plausible is
it?) that the  function $F$ constructed above is a
pseudorandom function}
:
\begin{equation}
\label{eq:Syst}
y_{0}=F_0(z), y_{0}\oplus y_1=F_1(z+1),\ldots,y_{s-2}\oplus y_{s-1}=F_{s-1}(z+s-1)
\end{equation}
To find $z$ the attacker may try to solve any of these equations%
; he could
do it 
with a negligible advantage, since $F_i$ is one-way. 
Of course, the attacker may
try to express $z+i$ as a collection of ANF's 
$\delta_0(z+i),\ldots,\delta_{n-1}(z+i)$
in the variables $\chi_0=\delta_0(z),\ldots,\chi_{n-1}=\delta_{n-1}(z)$, then substitute
these ANF's 
for the variables into  the ANF's 
that define mappings
$F_i$, to obtain an overdefined system \eqref{eq:Syst} in unknowns 
$\chi_0,\ldots,\chi_{n-1}$. However, the known formula (see e.g. \cite{Alp}
and
fix an obvious misprint there)
\begin{equation}
\label{eq:sumBool}   
\delta_j(z+i)\equiv \chi_j+\delta_j(i)+\sum_{r=0}^{j-1}\delta_r(i)\cdot\chi_r
\prod_{t=r+1}^{j-1}(\delta_t(i)+\chi_t)\pmod
2;
\end{equation}
implies that the number of monomials in the equations of the obtained system
will be, generally speaking, exponential in $n$; to say nothing of that
the number of operations
to make these substitutions and then to collect similar terms is also exponential
in $n$, unless the degree of all ANF's 
that define all $F_i$ is
bounded by a constant (the latter is not a case according to our
assumptions). 

Finally, our assumption that the attacker knows all $F_i$ seems
to be too strong: It is more practical
to assume that he does not know $F_i$
in \ref{eq:Syst}, since given clock output (and/or clock state update)
functions as explicit compositions of
arithmetical and bitwise logical operators, `normally' it is infeasible to express 
these functions in the Boolean form \ref{Bool}:
Corresponding ANF's
`as a rule' are sums of exponential in $n$
number of monomials, cf. \eqref{eq:sumBool}. 
Moreover, if these clock output functions $F_i$ and/or clock state update functions
$f_i$
are determined by  a key-dependent control sequence 
(say,  which is produced by a generator with unknown initial state), see Section \ref{sec:Constr},
then the explicit forms of the mentioned compositions are also unknown.
So in general an attacker has to find an initial state $u_0$ having only
a segment $z_j,z_{j+1},\ldots$ of the output sequence formed according
to the rule \eqref{eq:cntdpd}, where both $f_i$ and $F_i$ are not known
to him. An `algebraic' way to do this by guessing $f_i$ and $F_i$ and solving corresponding systems
of equations  seems to be hopeless in view of \ref{cor:Num} and the above discussion. The results
of preceding sections\footnote{as well as computer experiments: Output
sequences of explicit generators of the kind considered in the paper passed
both DIEHARD and NIST test suites} give us reasons to conjecture that under common tests the sequence
$z_j,z_{j+1},\ldots$ behaves like a random one, so `statistical' methods
of breaking such (reasonably designed) ciphers seem to be ineffective as well.


%

\bibliographystyle{plain}
\bibliography{mabit_05}

\begin{thebibliography}{10}

\bibitem{Alp}
R.~C. Alperin.
\newblock $p$-adic binomial coefficients $\bmod p$.
\newblock {\em The Amer. Math. Month.}, 92(8):576--578, 1985.

\bibitem{abc-v2}
V.~Anashin, A.~Bogdanov, and I.~Kizhvatov.
\newblock {ABC}: {A} {N}ew {F}ast {F}lexible {S}tream {C}ipher, {V}ersion 2.
\newblock Available from \url{http://crypto.rsuh.ru/papers/abc-spec-v2.pdf},
  2005.

\bibitem{me-1}
V.~S. Anashin.
\newblock Uniformly distributed sequences of $p$-adic integers.
\newblock {\em Mathematical Notes}, 55(2):109--133, 1994.

\bibitem{me-exp}
V.~S. Anashin.
\newblock Uniformly distributed sequences in computer algebra, or how to
  constuct program generators of random numbers.
\newblock {\em J. Math. Sci.}, 89(4):1355--1390, 1998.

\bibitem{me-2}
V.~S. Anashin.
\newblock Uniformly distributed sequences of $p$-adic integers, {II}.
\newblock {\em Discrete Math. Appl.}, 12(6):527--590, 2002.
\newblock A preprint available from \url{http://arXiv.org/math.NT/0209407}.

\bibitem{me-Kol}
V.~S. Anashin.
\newblock On finite pseudorandom sequences.
\newblock In {\em Kolmogorov and contemporary mathematics.}, pages 382--383,
  Moscow, June 2003. {Russian Academy of Sciences, Moscow State University}.
\newblock Abstracts of the Int'l Conference.

\bibitem{me-04}
V.~S. Anashin.
\newblock Pseudorandom number generation by $p$-adic ergodic transformations.
\newblock Available from \url{http://arxiv.org/abs/cs.CR/0401030}, January
  2004.

\bibitem{me-04a}
V.~S. Anashin.
\newblock Pseudorandom number generation by $p$-adic ergodic transformations:
  {A}n addendum.
\newblock Available from \url{http://arxiv.org/abs/cs.CR/0402060}, February
  2004.

\bibitem{me-conf}
Vladimir Anashin.
\newblock Uniformly distributed sequences over $p$-adic integers.
\newblock In I.~Shparlinsky A.~J. van~der Poorten and H.~G. Zimmer, editors,
  {\em Number theoretic and algebraic methods in computer science. Proceedings
  of the Int'l Conference (Moscow, June--July, 1993)}, pages 1--18. World
  Scientific, 1995.

\bibitem{abc_per}
Vladimir Anashin, Andrey Bogdanov, and Ilya Kizhvatov.
\newblock Increasing the {ABC} {S}tream {C}ipher {P}eriod.
\newblock Technical report, {ECRYPT}, July 2005.
\newblock \url{http://www.ecrypt.eu.org/stream/papersdir/050.pdf}.

\bibitem{BFS}
M.~Bardet, J.-C. Faug\`ere, and B.~Salvy.
\newblock Complexity of {G}r\"obner basis computation for semi-regular
  overdetermined sequences over {$\mathbb F_2$} with solutions in {$\mathbb
  F_2$}.
\newblock Available from \url{http://www.inria.fr/rrrt/rr-5049.html}, 2004.

\bibitem{XL}
N.~Courtois, A.~Klimov, J.~Patarin, and A.~Shamir.
\newblock Efficient algorithms for solving overdefined systems of multivariate
  polynomial equations.
\newblock In {\em Eurocrypt 2000}, volume 1807 of {\em Lect. Notes Comp. Sci.},
  pages 392--407. Springer-Verlag, 2000.

\bibitem{GJ}
M.~R. Garey and D.~S. Johnson.
\newblock {\em Computers and Intractability: A Guide to the Theory of
  {$NP$}-completeness}.
\newblock W.H. Freeman and Co., 1979.

\bibitem{Kl-Gor}
A.~Klapper and M.~Goresky.
\newblock Feedback shift registers, $2$-adic span, and combiners with memory.
\newblock {\em J. Cryptology}, 10:111--147, 1997.

\bibitem{KlSh:3}
A.~Klimov and A.Shamir.
\newblock New cryptographic primitives based on multiword {T}-functions.
\newblock In Bimal Roy and Willi Meier, editors, {\em Fast Software Encryption:
  11th International Workshop, FSE 2004, Delhi, India, February 5-7, 2004.
  Revised Papers}, pages 1 -- 15. Springer-Verlag GmbH, 2004.

\bibitem{KlSh-2}
A.~Klimov and A.~Shamir.
\newblock Cryptographic applications of {T}-functions.
\newblock In {\em Selected Areas in Cryptography -2003}, 2003.

\bibitem{KlSh}
A.~Klimov and A.~Shamir.
\newblock A new class of invertible mappings.
\newblock In B.S.Kaliski~Jr.et al., editor, {\em Cryptographic Hardware and
  Embedded Systems 2002}, volume 2523 of {\em Lect. Notes in Comp. Sci}, pages
  470--483. Springer-Verlag, 2003.

\bibitem{Knuth}
D.~Knuth.
\newblock {\em The {A}rt of {C}omputer {P}rogramming}, volume~2.
\newblock {A}ddison-{W}esley, {T}hird edition, 1998.

\bibitem{Kot}
L.~Kotomina.
\newblock Fast nonlinear congruential generators.
\newblock Diploma Thesis, Russian State University for the Humanities, Moscow,
  1999.
\newblock (in Russian).

\bibitem{KN}
L.~Kuipers and H.~Niederreiter.
\newblock {\em Uniform Distribution of Sequences}.
\newblock John Wiley \& Sons, N.Y. etc., 1974.

\bibitem{Lar}
M.~V. Larin.
\newblock Transitive polynomial transformations of residue class rings.
\newblock {\em Discrete Mathematics and Applications}, 12(2):141--154, 2002.

\bibitem{LN}
Hans Lausch and Wilfried N\"obauer.
\newblock {\em Algebra of Polynomials}.
\newblock North-Holl. Publ. Co, American Elsevier Publ. Co, 1973.

\bibitem{ShTs}
A.~Shamir and B.~Tsaban.
\newblock Guaranteeing the diversity of number generators.
\newblock {\em Information and Computation}, 171:350--363, 2001.
\newblock Available from \url{http: //arXiv.org/ abs/ cs.CR/ 0112014}.

\bibitem{Yb}
S.~V. Yablonsky.
\newblock Basic notions of cybernetics.
\newblock In {\em Problems of Cybernetics}. Fizmatgiz, 1959.
\newblock (in Russian).

\end{thebibliography}
\end{document}